\documentclass[twocolumn,showpacs,amsmath,amssymb,prl]{revtex4}
\usepackage{graphicx}
\usepackage{dcolumn}
\usepackage{bm}
\begin{document}
\title{
Localization in a quantum spin Hall system
}
\author{Masaru Onoda$^{1,4}$}
\email{m.onoda@aist.go.jp}
\author{Yshai Avishai$^{2,3}$}
\email{yshai@bgu.ac.il}
\author{Naoto Nagaosa$^{1,3,4}$}
\email{nagaosa@appi.t.u-tokyo.ac.jp}
\affiliation{
$^1$Correlated Electron Research Center (CERC),
National Institute of Advanced Industrial Science and Technology (AIST),
Tsukuba Central 4, Tsukuba 305-8562, Japan\\
$^2$Department of Physics, Ben-Gurion University 
, Beer-Sheva 84105 Israel \\
$^3$Department of Applied Physics, University of Tokyo, Bunkyo-ku, Tokyo 113-8656, Japan\\
$^4$CREST, Japan Science and Technology Corporation (JST), Saitama, 332-0012, Japan\\
}
\begin{abstract}
Localization problem of electronic states in a two-dimensional 
 quantum spin Hall system (QSH - a symplectic model with a non-trivial 
topological structure) 
is studied by the transfer matrix method. 
The phase diagram in the plane of energy and disorder strength is exposed, 
and demonstrates "levitation" and
"pair-annihilation" of the domains of extended states analogous to that of 
the integer quantum Hall system. The critical exponent $\nu$ for the 
divergence of the localization length is estimated as $\nu \cong 1.6$ 
which is distinct from both exponents pertaining to 
 the conventional symplectic and the unitary quantum Hall systems. This strongly
suggests a different universality class related to the non-trivial topology of
the QSH system. 
\end{abstract}
\maketitle

The spin Hall effect (SHE) in semiconductors, namely,
generation of spin current  perpendicular to an 
applied electric field, has recently attracted 
intensive attention following its theoretical proposals \cite{MNZ,MNZ1,Sinova}.  
Recent experiments observed an accumulated spin polarization near 
the edges of a sample in two-dimensional n-type \cite{Kato} and 
p-type \cite{Wunderlich} quantum wells of GaAs under an applied electric field. 
There still remain some controversial issues about the origin of 
these accumulated spins \cite{Halperin}, i.e., 
whether it is due to extrinsic \cite{Dyakonov,Hirsch,Zhang} 
or intrinsic \cite{MNZ,Sinova} origins. 
However, the fundamental importance of the SHE is 
the underlying topological structure of Bloch wavefunctions in 
 systems with time-reversal (TR) symmetry. The concept of SU(2) non-Abelian Berry phase 
plays a crucial role in the intrinsic SHE \cite{MNZ,MNZ1}, 
which is a generalization of the anomalous Hall effect in metallic ferromagnets driven 
by the U(1) Berry phase \cite{OnodaNagaosa}. 
These topological aspects led to the concept 
of spin Hall insulator (SHI) \cite{SHI}, which is a band insulator showing nonzero 
spin Hall conductivity. In these models \cite{SHI}, 
edge modes are shown to be gapful in generic cases \cite{OnodaSHI}, 
in sharp contrast with quantum Hall systems (see also ref.~\cite{Bernevig1}
where another proposal for the quantum spin Hall (QSH) system is discussed). 

Originally, zero gap semiconductors such as 
HgTe, $\alpha$-Sn and narrow gap semiconductors such as PbTe were listed 
as candidates for SHI \cite{SHI}. 
Later on, a model for QSH effect in graphene 
has been proposed \cite{Kane1,Kane2}, which is shown to be 
topologically distinct from the previous ones since 
it is characterized by an odd number $\mathrm{Z}_2$ of pairs of  
gapless edge modes for a semi-infinite system. 
In that case, Kramers theorem for systems with TR symmetry 
prevents the hybridization of edge modes 
with opposite chiralities \cite{Kane2,Xu,helicaledge}. 
The number $\mathrm{Z}_2$ so defined is claimed 
to be related to the mod 2 index of the real Dirac operator
and the latter is then explicitly constructed from 
the Bloch wavefunctions and used to distinguish 
the QSH phase from a simple SHI.
Note however that spin is not a conserved quantity 
in the presence of spin-orbit interaction, and one cannot define a
topological invariant of the first Chern class 
for the corresponding SU(2) gauge field \cite{MNZ1}. 
(There is one proposal for the conserved spin current \cite{Cons}
which applies only for spatially homogeneous systems.)
Therefore, spin Hall conductance is not related to 
a topological integer such as the U(1) Chern number $C_{\mathrm{U}(1)}$
for the charge quantum Hall effect (QHE) \cite{TKNN}.
In this context, a recent work introduces 
a spin Chern number $C_{sc}$ (and a Chern number {\it matrix}),
and investigates its stability toward disorder 
and its relationship to edge modes \cite{Sheng1,Sheng2,Qi}. 
An important consequence drawn therein
is that even though there is no conserved spin current, one can still define 
a conserved topological number associated with twisted (spin-dependent) boundary 
conditions \cite{Niu}. 
Here the analogy with the QHE is rather appealing.

A question of paramount interest is how the occurrence of 
these new topological quantities affect the localization/delocalization 
properties of wave functions in the presence of disorder. 
Although the standard symplectic ensemble exhibits 
a metal-insulator transition in two-dimensions, here we encounter a 
symplectic system with a non-trivial topological structure.
It is already known for the quantum Hall systems, i.e., 
the unitary case, that $C_{\mathrm{U}(1)}$ \cite{TKNN} 
associated with the conserved charge current protects 
the isolated extended states at some special energies \cite{Prui}, 
although the naive renormalization group (RG) study of the 
unitary universality class dictates the localization 
for any finite disorder strength in two-dimensions \cite{Lee}. 
Since $C_{\mathrm{U}(1)}$ is defined as a response to  twisted boundary conditions, 
existence of extended states is necessary for having a nonzero $C_{\mathrm{U}(1)}$.
This topological integer cannot change continuously.  
Rather, it jumps when two extended states 
with opposite Chern numbers {\it annihilate} each other. The pertinent field 
theory is the nonlinear-$\sigma$ model with a topological term, which 
results in two-parameter scaling RG equations \cite{Prui}. 
Assuming that $\mathrm{Z}_2$ classification or $C_{sc}$ is 
also associated with extended states, 
we expect a similar scenario for the localization problem 
in the symplectic model belonging to the non-trivial $\mathrm{Z}_2$ 
classification and/or with nonzero $C_{sc}$.  

Thus, in this work, we investigate the localization problem for a symplectic model 
with a nontrivial topological structure, i.e., QSH system 
with odd $\mathrm{Z}_2$ and $C_{sc}=2$. 
Using the transfer matrix method, the phase diagram in the plane of disorder strength 
and energy is revealed, which manifests the features of {\it levitation} and 
{\it pair annihilation} of extended states similar to the unitary QHE case, 
albeit with finite energy width of the extended region. Finite size scaling analysis 
of the localization/delocalization transition yields an exponent $\nu \cong 1.6$ 
for the divergence of the localization length, which is distinct from both 
that of the symplectic ($\nu\cong 2.73$) \cite{Ohtsuki}
and that of the unitary QHE ($\nu\cong 2.33$) \cite{IQHE}
universality classes. 
This strongly suggests a new universality class of the
localization/delocalization transition in the symplectic class, 
due to the existence of a non-trivial topological structure.

The model Hamiltonian we study is a generalization of 
that proposed for graphene by Kane and Mele \cite{Kane1,Kane2}:
\begin{eqnarray}
H &=& \sum_{\langle ij\rangle} c^\dagger_i\left[ t 
+i\lambda_{\mathrm{R}}
(\bm{\sigma} \times {\hat{\bm{d}}_{ij}} )_z\right]c_j 
\nonumber\\
&+&\sum_{\langle ij\rangle'}c^\dagger_i(
t'+ i\lambda_{\mathrm{SO}}\nu_{ij}\sigma_z) c_j 
\nonumber  \\
&+& \sum_i 
c^\dagger_i(
\lambda_v \eta_i  
+ w_i 
+ h_x \sigma_x)c_i.
\end{eqnarray}
Here $c_i^{(\dagger)}$ is the spinor annihilation (creation) operator,
$\bm{\sigma}$ is the set of the Pauli matrices,
$t$ is the conventional hopping energy between nearest neighbor sites $\langle ij\rangle$
(we use $t=1$ as the unit of energy),
$\lambda_{\mathrm{R}}$ is the Rashba spin-orbit interaction strength
and $\hat{\bm{d}}_{ij}$ is the unit vector connecting $\langle ij\rangle$.
The constant $t'$ is the conventional hopping energy between 
second nearest neighbor sites $\langle ij\rangle'$.
Here it is introduced just for assuring stability of the numerical analysis 
by the transfer matrix method and is fixed at the small value 0.01.
Moreover, $\lambda_{\mathrm{SO}}$ represents the spin-orbit interaction strength with 
$\nu_{ij} = (2/\sqrt{3}) ( {\hat{\bm{d}}}_1 \times {\hat{\bm{d}}}_2 ) = \pm 1$, 
where ${\hat{\bm{d}}}_1$ and ${\hat{\bm{d}}}_2$
are unit vectors along the two bonds connecting $\langle ij\rangle'$,
while $\lambda_v$ is the alternation of the site energies 
between the A and B sublattices ($\eta_i=\pm1$). 
The random potential $w_i$ is uniformly distributed between $-W/2$ and $W/2$. 
Finally, the last term represents the magnetic field along the $x$-direction, 
which breaks TR symmetry. 
The phase diagram of this model without the random potential $w_{i}$
and the magnetic field $h_x$ has 
been already displayed in the inset of Fig.~1 in ref.~\cite{Kane1}.
In the plane $\lambda_{v}/\lambda_{\mathrm{SO}}$-$\lambda_{\mathrm{R}}/\lambda_{\mathrm{SO}}$, 
there is a finite domain of QSH state with $Z_2=1$, $C_{sc}=2$, 
bounded by a curve where the gap closes. 
Outside this domain, the gap opens up again and the system becomes 
the usual SHI with $Z_2=0$, $C_{sc}=0$. 
Note that for $\lambda_{\mathrm{R}} =0$, the system is decoupled into two independent
unitary subsystems  \cite{Haldane,OnodaLoc}. 
For $\lambda_v < \lambda_{v}^c$, each unitary model has $C_{\mathrm{U}(1)}$'s 
of opposite signs for the valence and conduction bands, 
and the way of distribution of $C_{\mathrm{U}(1)}$'s is opposite for each unitary model.
Thus the total $C_{\mathrm{U}(1)}$ of valence bands is zero,
but $C_{sc}=2$.
In the case of $\lambda_v > \lambda_{v}^c$, $C_{\mathrm{U}(1)}$'s of each band vanishes,
and $C_{sc}$ is also zero.
For finite $\lambda_{\mathrm{R}}$, these two unitary models are hybridized 
and become a symplectic model.
$C_{sc}$ of this hybridized system is quantized as $C_{sc}=2$ 
in the domain of the QSH state,
while it vanishes in the domain of the usual SHI.
With a finite magnetic field $h_x$, similar hybridization occurs,
but this breaks TR symmetry, and the model belongs to 
the topologically-trivial unitary class
where $C_{\mathrm{U}(1)}$ of each band is zero.
Interestingly, in the clean limit,
$C_{sc}$ of this unitary model is still quantized,
i.e, $C_{sc} =2$, while each pair of edge states opens 
a gap due to the hybridization by $h_x$.
(Note that $C_{sc}$ is well-defined even in TR breaking systems.)
  
The localization length $\lambda_{M}(W,E)$ of a quasi 
one-dimensional tube of $M$-site circumference
is calculated at  energy $E$ and disorder strength 
$W$ by the transfer matrix method~\cite{ZPB53_0000001_83}.
The $M$-dependence of the renormalized localization length 
$\Lambda_{M}(W,E)=\lambda_{M}(W,E)/M$ determines the 
localization/delocalization properties of the wavefunctions at energy $E$.

\begin{figure*}[hbt]
\includegraphics[scale=0.54]{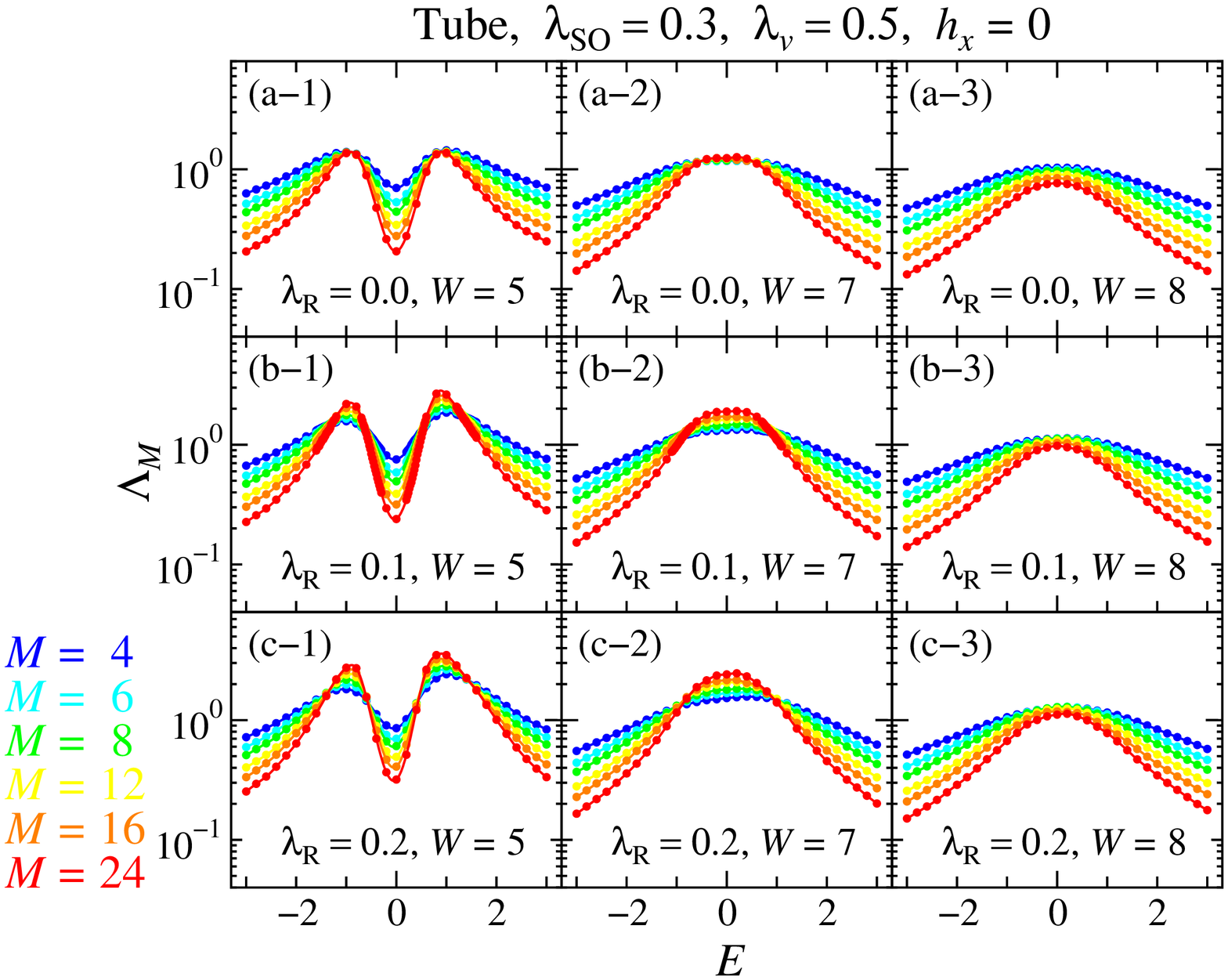}
\includegraphics[scale=0.44]{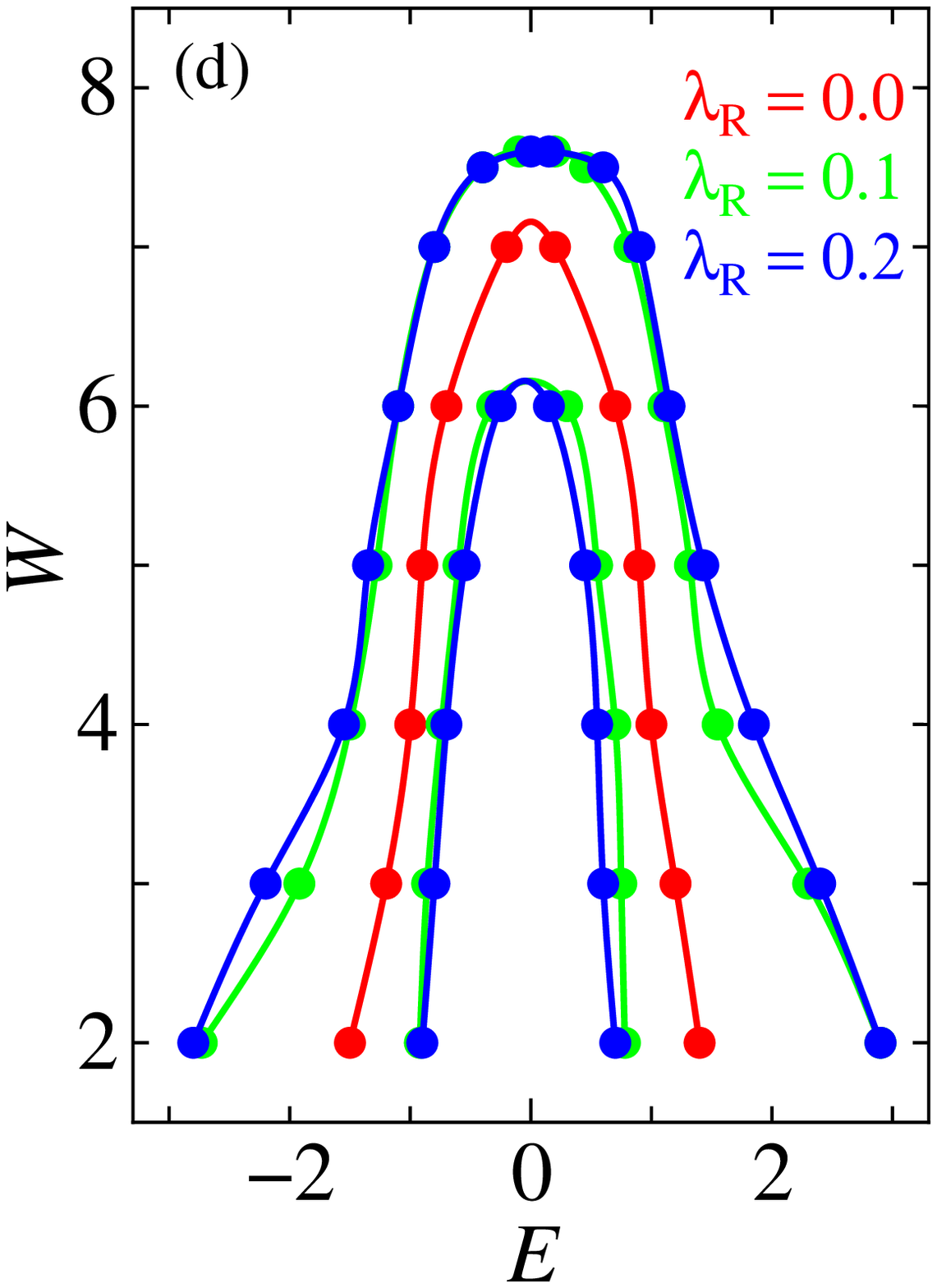}
\caption{
Renormalized localization length $\Lambda_{M}(W,E)=\lambda_{M}(W,E)/M$
for QSH with $\lambda_{\mathrm{R}}=0$ (a-1,2,3), 
$\lambda_{\mathrm{R}}=0.1$ (b-1,2,3) and $\lambda_{\mathrm{R}}=0.2$ (c-1,2,3).
The disorder strength $W$ is increased from left to right as
$W=5.0$ (a,b,c-1), $W=7.0$ (a,b,c-2) and $W=8.0$ (a,b,c-3). 
The other parameters are fixed as $\lambda_{\mathrm{SO}} =0.3$, 
$\lambda_v=0.5$ and $h_x=0$.
(d) A localization/delocalization phase diagram obtained in the plane of 
energy $E$ and disorder strength $W$.  
The red curve is the energy of the isolated extended states for 
$\lambda_{\mathrm{R}} = 0$, while the green and blue curves are the boundary of 
the energy region of the extended states for $\lambda_{\mathrm{R}}=0.1$ and 
$\lambda_{\mathrm{R}}=0.2$, respectively.
} 
\label{fig:lambda_QSH}
\end{figure*}  
Figures~\ref{fig:lambda_QSH}(a-1)-(c-3) display  $\Lambda_{M}(W,E)$ 
up to $M=24$ for several values of $W$ as functions of $E$
for disordered QSH system at $h_x=0$ (symplectic ensemble).
The case $\lambda_R=0$ has been studied in the 
context of quantized anomalous Hall effect \cite{OnodaLoc}, 
and the two isolated extended states are identified by the $M$-independent 
$\Lambda_{M}(W,E)$. There are two energies at which extended states show up. 
As they merge together the extended states disappear. 
This is consistent with the scenario of levitation and pair annihilation
of the two $C_{\mathrm{U}(1)}$'s of opposite signs.
With finite $\lambda_R$, the isolated extended states turn
into finite energy region of extended states, the width of which
increases with $\lambda_R$. 
Note that these two regions of extended states approach 
as the disorder $W$ increases.
The gap in the density of states disappears already 
for $W$ larger than $\sim 3$, but still the two regions are separated.
When $W$ is further increased, these two energy regions
of extended states merge into one region, and eventually disappear.
We have also checked that, in the ribbon geometry,
there appear extended gapless edge states even when there are no 
extended bulk states in the middle energy region.
Based on these results, we draw in Fig.~\ref{fig:lambda_QSH}(d) a phase diagram
depicting the location of extended states in the  $E$-$W$ plane.
The red curve for $\lambda_R=0$ represents the trajectory of 
the isolated extended states in the unitary (QHE) case,
while the phase boundaries between the localized
and extended states are given by green curves for 
$\lambda_R=0.1$ and by blue ones for $\lambda_R=0.2$, respectively.

\begin{figure}[hbt]
\includegraphics[scale=0.42]{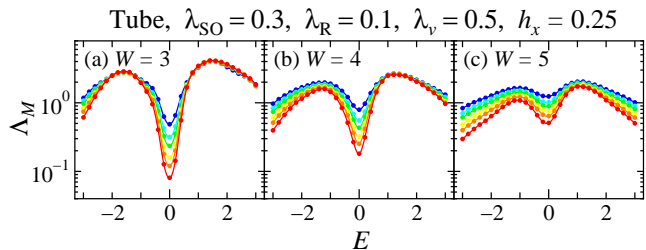}
\caption{
Renormalized localization length $\Lambda_{M}(W,E)=\lambda_{M}(W,E)/M$
for the unitary model with $h_x=0.25$ at (a) $W=3.0$, (b) $W=4.0$, (c) $W=5.0$. 
The other parameters are 
$\lambda_{\mathrm{SO}} = 0.3$, $\lambda_{\mathrm{R}}=0.1$ and $\lambda_v=0.5$. 
All the states are already localized 
since the U(1) Chern numbers for up and down spin bands cancel out.
} 
\label{fig:lambda_u-SHI}
\end{figure}
\begin{figure}[hbt]
\includegraphics[scale=0.42]{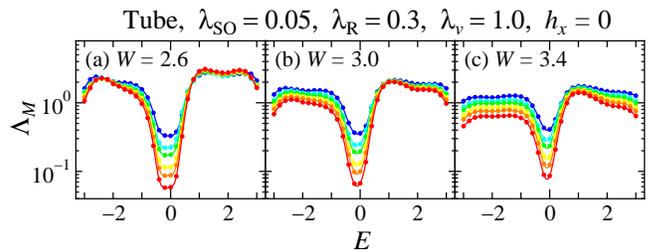}
\caption{
Renormalized localization length $\Lambda_{M}(W,E)=\lambda_{M}(W,E)/M$
for SHI with $\lambda_{\mathrm{SO}} = 0.05$, 
$\lambda_{\mathrm{R}}=0.3$ and $\lambda_v=1.0$ 
at (a) $W=2.6$, (b) $W=3.0$ and (c) $W=3.4$.
} 
\label{fig:lambda_s-SHI}
\end{figure}
Let us confront some other cases with those discussed in Fig.~\ref{fig:lambda_QSH}.
Figures~\ref{fig:lambda_u-SHI}(a)-(c) display the curves $\Lambda_M(W,E)$ for the unitary model 
with $h_x=0.25$ at (a) $W=3.0$, (b) $W=4.0$ and (c) $W=5.0$. In this case, the extended states 
have already disappeared since $C_{\mathrm{U}(1)}$ is zero 
for each of the split bands. Therefore the model is reduced to 
the trivial unitary class, where all states in two dimensions are localized 
with any finite amount of disorder \cite{Lee}. 
(It is also confirmed that the gapful edge states 
in the ribbon geometry are localized.)
However, it should be noted that $C_{sc}$ for this system is
quantized as $C_{sc}=2$ in the clean limit.
This means that, in TR breaking systems, finite $C_{sc}$ does not 
protect extended states and that protection would be 
closely related to the Kramers theorem.
One might think that the difference is due solely to symmetry, i.e., 
unitary {\it v.s} symplectic classes, and not to the topological
property of the QSH state. It is therefore important 
to compare with the simple symplectic model
which belongs to the trivial $\mathrm{Z}_2$ classification.
In our model, the SHI corresponds to this case, 
and Fig.~\ref{fig:lambda_s-SHI} shows $\Lambda_M(W,E)$ for $\lambda_{SO}=0.05$, $\lambda_R=0.3$,
$\lambda_v=1.0$ at (a) $W=2.6$, (b) $W=3.0$ and (c) $W=3.4$, respectively.
It is evident here that the extended states disappear with 
much weaker disorder strength, and the two energy regions of
extended states disappear without merging into a single one.

The above two cases, i.e., Figs.~\ref{fig:lambda_u-SHI} and \ref{fig:lambda_s-SHI} 
strongly suggest that the localization behavior of QSH system 
in Fig.~\ref{fig:lambda_QSH}
is deeply influenced by the nontrivial topological aspect, and
is distinguished from that of the usual symplectic class.
In order to substantiate this expectation, we have studied the 
critical property of the localization/delocalization 
transition of the QSH system. 

\begin{figure}[hbt]
\includegraphics[scale=0.4]{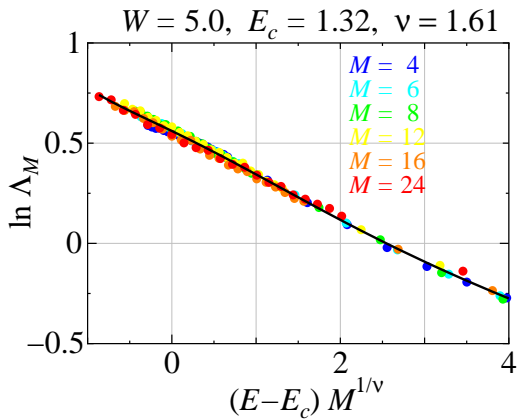}
\caption{
Scaling plot of the renormalized localization length
$\Lambda_{M}(W,E)$ (see raw data in Fig.~\ref{fig:lambda_QSH}(b-1)) 
as a function of $|E- E_c|M^{1/\nu}$
for different energies $E$ and $M \le 24$ and for $W=5$. 
The critical energy and exponent are estimated as
$E_c= 1.32 \pm 0.005$, $\nu=1.61 \pm 0.10$, respectively.
} 
\label{fig:scaling}
\end{figure}
Figure~\ref{fig:scaling} summarizes the scaling analysis by displaying 
$\Lambda_M(W,E) = f(|E-E_c| M^{1/\nu})$ at $W=5.0$
with $\nu$ being the critical exponent for the divergence of the
localization length. Data for various $E$ and $M$ (up to $M=24$) are 
included and their collapse on a single curve indicates a reasonable
 one parameter scaling behavior, which simultaneously determines
$\nu = 1.61 \pm 0.10$.
We have also studied transition at higher disorder
 $W=7.0$, and found $\nu = 1.61 \pm 0.10$.
This exponent should be confronted with that
of the standard symplectic universality class $\nu\cong 2.73$
\cite{Ohtsuki}, and that of the unitary model at strong magnetic field 
(with finite $C_{\mathrm{U}(1)}$) $\nu\cong 2.33$ \cite{IQHE}.
The result $\nu\cong 1.6$ obtained here is clearly distinct from both of these
values, and suggests a new universality class for the 
symplectic ensemble with a non-trivial topological structure.
Intuitively,  it is anticipated that the localization problem should be
influenced by the finite spin Chern number.  The 
construction of an effective field theory for this class
is left for future investigations.

In summary, we have studied the localization/delocalization 
problem of the QSH system 
(a symplectic ensemble with non-trivial topological property) 
by the transfer matrix method.  The phase diagram 
shows levitation and pair annihilation of the two 
energy regions of extended states, analogous to that of
the unitary model with finite Chern number 
(the integer quantum Hall effect). The critical
exponent $\nu$ for the localization/delocalization 
transition is estimated as $\nu \cong 1.6$, which
is distinct from that of 
the standard symplectic universality class ($\nu\cong 2.73$) 
and that of the unitary class with non-zero Chern number ($\nu\cong 2.33$).
This strongly suggests that the QSH system belongs to a new
universality class characterized by a topological index
such as the spin Chern number matrix or the $\mathrm{Z}_2$ index.

The authors would like to thank C.~L.~Kane, F.~D.~M.Haldane, 
S.~C.~Zhang and T.~Ohtsuki for fruitful discussions.
This work is financially supported by NAREGI Grant,
Grant-in-Aids from the Ministry of Education,
Culture, Sports, Science and Technology of Japan and 
JSPS long term program (YA).

\end{document}